\documentclass[usenatbib]{tjaa}

\usepackage[utf8]{inputenc}
\usepackage{lipsum}
\newsavebox\verbbox

\usepackage[T1]{fontenc}
\usepackage{amsmath}		% Advanced maths commands
\usepackage{amssymb}		% Extra maths symbols
\usepackage[pdftex]{graphicx}	% Including figure files
\usepackage[utf8]{inputenc}
\usepackage[english]{babel}

\usepackage{lipsum}
\usepackage{gensymb}
\usepackage{textcomp}

\title[HAT-P-16b ve TrES-3b]{Geçiş Zamanları Değişimi Yöntemiyle HAT-P-16b ve TrES-3b Ötegezegenlerinin İncelenmesi}%%%%TJAA-OZEL:TITLE%

\author[]{
Y. Aladağ,$^{1}$\thanks{E-mail: aladagyasmin@gmail.com}\orcid{0000-0002-6107-579X}
A. Akyüz,$^{1,3}$
Ö. Baştürk,$^{4}$
N. Aksaker,$^{1,2}$
E.M. Esmer,$^{4}$
S. Yalçınkaya,$^{4}$
\\
$^1$Uzay Bilimleri ve Güneş Enerjisi Araştırma ve Uygulama Merkezi (UZAYMER), Çukurova Üniversitesi, 01330, Adana, Turkey\\
$^2$Adana Organize Sanayi Bölgesi Teknik Bilimler Meslek Yüksekokulu, Çukurova Üniversitesi, 01330, Adana, Turkey\\
$^3$Fizik Bölümü, Çukurova Üniversitesi, 01330, Adana, Turkey\\
$^4$Astronomi ve Uzay Bilimleri Bölümü, Ankara Üniversitesi, Ankara, Turkey\\
}
\date{Accepted: XXX. Revised: YYY. Received: ZZZ.}
\renewcommand{\pubyear}{0000}
\renewcommand{\volume}{0}
\renewcommand{\issue}{0}
\begin{document}
% Don't change these 3 lines
\label{firstpage}
\pagerange{\pageref{firstpage}--\pageref{lastpage}}
\maketitle{M00-0000}

\begin{abstract}
In this study, the transit observations of HAT-P-16b and TrES-3b exoplanets were carried out at ÇÜ UZAYMER Observatory with a 50 cm Ritchey Chretien type telescope. By analyzing the light curves of both exoplanets, system parameters have been obtained with acceptable models. Some of these parameters for HAT-P-16 systems were found to be: M$_{P}$=4.172$\pm$0.163 M$_{J}$, R$_{P}$=1.309$\pm$0.111 R$_{J}$, a/R$_{*}$=7.1922$\pm$0.0017 and b=0.1003$\pm$0.1533 which are consistent with the values given in \cite{2010ApJ...720.1118B}. Also, for the TrES-3 system, the same parameters were calculated as M$_{P}$=1.959$\pm$0.111 M$_{J}$, R$_{P}$=1.320$\pm$0.169 R$_{J}$, a/R$_{*}$=6.0656$\pm$ 0.4899 and b=0.7892$\pm$0.0700. These values are also consistent with the values given in \cite{2007ApJ...663L..37O}. No significant Transit Timing Variations (TTV) was found as a result of the Lomb-Scargle (LS) periodogram for HAT-P-16b by using the transit times obtained from the transit light curves (False Alarm Probabilities -- FAP) = \%96). However, a significant periodicity was found at 32.38 days and a secondary periodicity was found at 41.07 days were determined as a result of the LS periodogram obtained from the TrES-3b data. We interpret that  these periods with FAP of 3.41 \% and 1.88 \%, respectively, are close to the rotation period of the host star. Therefore, these periods could be due to the modulation of spot-induced light variations.

\ozet
Bu çalışmada, HAT-P-16b ve TrES-3b ötegezegenlerinin geçiş gözlemleri ÇÜ UZAYMER Gözlemevi’nde bulunan 50 cm (UT50) Ritchey Chretien tipi teleskobu ile gerçekleştirilmiştir. Her iki ötegezegenin ışık eğrileri analizleri yapılarak uygun modeller ile sistem parameterleri elde edilmiştir. HAT-P-16 için bu sistem parametrelerinden bazıları: M$_{P}$=4.172$\pm$0.163 M$_{J}$, R$_{P}$=1.309$\pm$0.111 R$_{J}$, a/R$_{*}$=7.1922$\pm$0.0017 ve b=0.1003$\pm$0.1533 olup \cite{2010ApJ...720.1118B} çalışmasında verilen değerler ile uyumludur. TrES-3 sistemi için bu parametreler: M$_{P}$=1.959$\pm$0.111 M$_{J}$, R$_{P}$=1.320$\pm$0.169 R$_{J}$, a/R$_{*}$=6.0656$\pm$ 0.4899 ve b=0.7892$\pm$0.0700 olup \cite{2007ApJ...663L..37O} çalışmasında sunulan değerler ile uyumludur. HAT-P-16b için geçiş ışık eğrileri kullanılarak oluşturulan Lomb-Scargle (LS) periyodogramı ile yapılan frekans analizi sonucunda anlamlı (Yalancı Alarm Olasılıkları -- FAP=\%96) bir Geçiş Zamanları Değişimi (TTV) bulgusuna rastlanmamıştır. Bununla birlikte TrES-3b ışık eğrilerinden elde edilen LS periyodogramında 32.38 günlük bir baskın dönem ve 41.07 günlük ikincil dönem tespit edilmiştir. FAP değerleri sırasıyla \% 3.41 ve \% 1.88 olan bu dönemliliklerin yıldızın dönme dönemine yakın olduğundan leke kaynaklı ışık değişimleri modülasyonlarından kaynaklanabileceğini değerlendirmekteyiz.
\end{abstract}


%%%%TJAA-OZEL:BILDIRI%
\section{Giriş} 
Güneş’ten başka bir yıldızın etrafında dolanan gezegenlere ötegezegen (exoplanet) adı verilmektedir. İlk ötegezegen 1992 yılında, bir atarca olan PSR 1257+12’nin çevresinde bulunmuştur \citep{1992Natur.355..145W}. Ayrıca çevresinde bir ötegezegen olduğu belirlenen ilk anakol yıldızı ise güneş benzeri bir yıldız olan 51 Pegasi’dir \citep{1995Natur.378..355M}. Bu keşiflerden sonra ötegezegen araştırmaları astronomi alanında çok çalışılan konular arasında yer almıştır. Teknolojik gelişmelerin katkısıyla günümüzde keşfedilen ötegezegen sayısı her geçen gün artış göstermektedir. Ötegezegen keşifleri farklı yöntemlerle yapılabilmektedir. Bugüne kadar sayıları 4311’i bulan ötegezegenlerden, 1’i astrometri (astrometry), 106’sı kütleçekimsel mercek (gravitational microlensing), 51’i doğrudan görüntüleme (direct imaging), 829’u dikine hız (radial velocity, RV), 3322’si geçiş (transit) yöntemiyle keşfedilmiştir\footnote{\url{https://exoplanets.nasa.gov/alien-worlds/ways-to-find-a-planet/}}. Yaygın olarak kullanılan geçiş yönteminde hem ötegezegen hem de barınak (host) yıldızın özellikleri hakkında bilgi edinilir. Geçiş yapan bir ötegezegen Kepleryan olduğu varsayılan bir yörüngede eşit zaman aralıklarında, dönemli olarak barınak yıldızının önünden geçer. Sistemde ikinci bir gezegen varsa yörüngeleri Kepleryan yörüngelere benzemez ve geçişler eşit aralıklı olmaz \citep{2005MNRAS.359..567A}, yani gözlenen dönem değişir. Geçiş zamanı değişimi (TTV; Transit Timing Variations) yöntemi, bilinen ötegezegenlerin geçiş zamanındaki değişikliklerin izlenmesine dayanır \citep{2009ApJ...701.1116N}. Gözlenen minimum ışık zamanı (O), hesaplanan minimum ışık zamanı (C) olmak üzere O-C diyagramları çift yıldızların dönem değişimlerini çalışmak için uzun zamandır kullanılmaktadır. Günümüzde ötegezegen araştırmalarında da O-C diyagramları yaygın olarak kullanılmaktadır. Hassas ölçümlerden elde edilmiş çok sayıda döneme ait geçiş ortası zamanları kullanılarak oluşturulan O-C diyagramları, TTV analizinin standart veri setini oluşturur. Geçiş zamanı değişimleri ilk olarak geçiş yöntemiyle aynı yıldız etrafında keşfedilen Kepler-9b ve Kepler-9c ötegezegenlerinin geçişlerinde gözlenmiştir \citep{2010Sci...330...51H}. Bu yöntemle keşfedilen ilk gezegen ise Kepler-19c dir \citep{2011ApJ...743..200B}. Yapılan TTV analizlerinde dönem bulabilmek için yaygın olarak Lomb-Scargle (LS) periyodogramları kullanılmaktadır. \cite{1976Ap&SS..39..447L} ve \cite{1982ApJ...263..835S} O-C diyagramında çevrimsel bir değişimin varlığını araştırmak için O-C artıklarından (residual) LS periyodogramları elde etmişlerdir. LS metodu, eş zaman aralıklı olmayan veride, zayıf periyodik sinyalleri bulmanın ve bunların önemini test etmede izlenen bir yoldur \citep{1989ApJ...338..277P}. 

TTV yöntemiyle HAT-P-16b ve TrES-3b ötegezegenlerinin geçiş zamanları değişimini analiz etmek üzere yapılan bu çalışma için UZAYMER Gözlemevi’nde bulunan UZAYMER Teleskobu (UT50) ile geçiş gözlemleri yapılmıştır. Sistem parametrelerinin belirlenmesinde UT50’den alınan geçiş verilerine ek olarak T100 (TUG; TUBİTAK Ulusal Gözlemevinde), IST60 (İstanbul Üniversitesi Gözlemevi Uygulama ve Araştırma Merkezine ait) ve ATA50 (Atatürk Üniversitesi Astrofizik Uygulama ve Araştırma Merkezinde) teleskoplarından veriler alınmıştır. Ayrıca TESS (Transiting Exoplanet Survey Satellite, \citealp{2009AAS...21340301R}) verileri de çalışmamızda kullanılmıştır. TTV analizlerinde sistem parametrelerinin belirlenmesinde kullanılan ışık eğrilerinden ölçülen geçiş ortası zamanlarının yanı sıra Exoplanet Transit Database (ETD, \citealp{2010NewA...15..297P}) ve literatürde yayınlanan geçiş ışık eğrilerinin yeniden modellenmesiyle ölçülen geçiş ortası zamanları kullanılarak O-C ışık eğrileri oluşturulmuştur. TTV analizlerinde dönem bulabilmek için LS periyodogramlarından yararlanılmıştır.

\textbf{HAT-P-16b} \\
HAT-P-16b ötegezegeni 2009 yılında yapılan gözlemlerle, geçiş yöntemi kullanılarak keşfedilmiştir \citep{2010ApJ...720.1118B}. Bu ötegezegen, parlaklığı $m_V$=10.8 kadir değerine sahip, F8 tayfsal tipteki barınak yıldızının etrafında ${\sim}$2.8 günlük yörünge döneminde dolanmaktadır ve yıldızının önünden geçiş süresi 183.7$\pm$1.9 dakikadır. Gezegen kütlesi M$_{P}{\sim}$4M$_{J}$ ve yarıçapı R$_{P}{\sim}$1.2 R$_{J}$ olarak belirlenmiştir.
\cite{2011A&A...533A.113M}'nın HAT-P-16b ile ilgili yaptıkları çalışmada, Haute-Provence Gözlemevi'nde 1.93 m teleskop üzerinde bağlı SOPHIE tayfçekeri ile yapılan tayfsal gözlemlerden elde edilen Rossiter-McLaughlin Etkisi \citep{2005ApJ...622.1118O}'nin (RME) büyüklük ve şekline dayanarak, HAT-P-16b ötegezegeninin yörünge düzlemiyle barınak yıldızın dönme ekseni arasındaki açının gökyüzü düzlemine izdüşüm açısını (sky-projected spin-orbit alignment angle) $\lambda = -10^{\circ}\pm16^{\circ}$ ve yıldızın dönme hızını $\upsilon.sini_{*}$=3.9$\pm$0.8 km.s$^{-1}$ olarak hesaplamışlardır. 
\cite{2012MNRAS.422.3151H}, yıldızlarına yakın gezegenlerin yörüngesel evrimlerini incelemek için bilinen geçiş ötegezegenlerinin dikine hız ölçümlerini analiz etmiştir ve HAT-P-16b’ nin dış merkezliğini e=0.034$\pm$0.003 olarak belirlemişlerdir. Sıfır olmayan bu dış merkezlik değeri, yörüngeyi tedirgin eden, gözlenmemiş ancak sisteme kütle çekimle bağlı olabilecek bir gezegenin varlığına bağlanabilir. HAT-P-16b ile ilgili başka bir çalışmada \cite{2013A&A...557A..30C}, eşzamanlı olarak iki orta sınıf teleskoptan odak dışı gözlem (defocusing) tekniğiyle \citep{2009MNRAS.394..272S, 2015ASPC..496..370B} HAT-P-16b’nin geçiş gözlemlerini yapmıştır. TTV analizi sonucunda sistemde 3. cismin varlığına dair herhangi bir kanıt bulamamışlardır. Bu ötegezegen hakkında yapılan çalışmaların az olması, bu çalışmada seçtiğimiz kaynaklardan biri olmasının önemli bir nedenidir.

\textbf{TrES-3b}\\
TrES-3b ötegezegeni 2007 yılında yapılan gözlemlerle, geçiş yöntemi kullanılarak keşfedilmiştir \citep{2007ApJ...663L..37O}. Kaynak, parlaklığı $m_V$=12.4 kadir değerine sahip, G tayf türünden barınak yıldızının etrafında ${\sim}$1.3 gün yörünge döneminde dolanmaktadır ve geçiş süresi 77.9$\pm$1.9 \citep{2012PASP..124..212S} dakikadır. Gezegen M$_{P}{\sim}$1.9 M$_{J}$ kütleye ve R$_{P}{\sim}$1.3 R$_{J}$ yarıçapa sahiptir. Bu ötegezegenin geçiş ışık eğrisi ile ilgili en çarpıcı özellik, yüksek etki parametresi nedeniyle (b=0.83) geçiş sinyalinin {\textquotedblleft}V{\textquotedblright} şeklindeki yapısıdır. Bu yapı geçiş ortası zamanının hassas belirlenmesine ve bu nedenle geçiş zamanlarının değişiminin çalışılmasına yardımcı olmaktadır. TrES-3b ile ilgili yapılan diğer bir çalışmada \cite{2009ApJ...700.1078G}, TrES-3 sisteminde varsayımsal ikinci gezegen varlığını TTV yöntemiyle araştırmak için, Liverpool Teleskoptaki RISE cihazı ile yaptıkları gözlem verilerini ve önceki gözlem verilerini kullanmışlardır. İstatistiksel olarak anlamlı bir TTV sinyali belirlenememesine rağmen, TrES-3b geçişlerinden elde edilen O-C grafiğinden hareketle, sistemde varsayımsal ikinci bir gezegenin dönem oranının fonksiyonu olarak üst kütle sınırları belirlenmiştir. Ayrıca \cite{2011PASJ...63..301L}, SOAO (Sobaeksan Optical Astronomy Observatory) ve LOAO (Lemmon Optical Astronomy Observatory) gözlemevlerinden alınan gözlem verileriyle TrES-3b'nin geçiş ışık eğrilerinin analizlerini yapmışlardır. Çalışmada, TTV analizi sonucunda 1.97 gün dönemli frekans bulunarak bunun, yıldız lekesi gibi manyetik aktiviteye bağlı bir yüzey parlaklık düzensizliğinden kaynaklandığı belirtilmiştir. 

\cite{2017NewA...55...39P} TrES-3b ile ilgili yaptıkları çalışmada, Çanakkale Onsekiz Mart Üniversitesi Gözlemevi'nde (ÇOMUO) kurulu olan T122 ve T60 teleskopları ile T100 (TUG) teleskobunu kullanarak geçiş gözlemi verilerinin analizlerinden sistem parametrelerini güncellemiş ve TTV ölçümü yapmışlardır.  TTV analizinde TrES-3b için önemli bir sinyal belirlenmemiştir, 84.79$\pm$0.35 günlük bir TTV dönemine karşılık gelen 0.0154$\pm$0.0001 çevrim/dönem frekans bulunmuştur. \cite{2020AJ....160...47M}'nın TrES-3b ile ilgili yaptıkları çalışmada, 2012-2018 yılları arasında elde edilen 12 geçiş ışık eğrisi ile birlikte önceki çalışmalardan 71 geçiş ışık eğrisi verilerinin, geçiş ortası zamanlarını kullanarak yeni doğrusal efemerisler türetmiş ve TTV analizinde sistemde 3. cisim olasılığını güçlendiren sonuçlar elde etmişlerdir. Bununla birlikte, frekans analizi ile olası TTV’nin periyodik olma ihtimalinin belirlenen istatistiksel anlamlılık seviyesinin düşük olmasından dolayı sistemde ek cismin varlığının zayıf bir olasılık olduğunu değerlendirmişler ve daha çok gözlem için sistemin takibini önermişlerdir. TrES-3b’nin genelleştirilmiş LS periyodogramında istatistiksel olarak en anlamlı pik değerini 0.043867 çevrim/dönem (29.78 gün) bulmuşlardır. Yıldızın belirlenen dönme dönemi (27.43) bu döneme yakındır. Bu durum, daha önce \cite{2017A&A...608A..26M} tarafından manyetik etkinlik kökenli lekelerin zamanlamaya etkili olmayabileceği şeklindeki önerilerin de tartışmaya açık olduğunu göstermektedir. Zira bu dönemlilik, leke kaynaklı ışık eğrisi asimetrilerinin geçiş zamanı değişimlerine neden olmasından kaynaklanabilir. Sistem, olası TTV için defalarca çalışılmışsa da bu çalışmalar eksik, güncel olmayan veriler üzerinden farklı araştırmacıların farklı yöntemlerle yaptığı geçiş zamanı ölçümlerine dayanmaktadır. Bu ötegezegen, sistemde olası bir 3. cismin tedirginlik etkisinden kaynaklanabilecek, sıfırdan farklı bir dış merkezlilik değerine sahip olması ve {\textquotedblleft}V{\textquotedblright} şeklindeki geçiş sinyalinin bu olgunun tespitinde önemli bir avantaj oluşturacak olması nedeniyle seçilmiştir.

\section{Gözlemler ve Veri Analizi}
Bu çalışmada incelenen ötegezegenler için gözlem verileri Çukurova Üniversitesi Uzay Bilimleri ve Güneş Enerjisi Araştırma ve Uygulama Merkezi’nde (UZAYMER) konuşlandırılmış 50 cm ayna çaplı f/8 odak oranına sahip Ritchey Chretien (RC) tipi UZAYMER Teleskobu (UT50) ile alınmıştır. UZAYMER'in teknik altyapısı, astrometeorolojik ve gözlem şartları ayrıntılı olarak araştırılmıştır \citep{tjaa874400}.
Kaynaklarının UT50 gözlem bilgileri Tablo \ref{T:ut50_obs}’de gösterildiği gibidir. TTV ölçümünde UT50 geçiş gözlemi verilerine ek olarak, Tablo \ref{T:obs}’de optik sınıfı RC olan, 100 cm ayna çaplı f/10 odak oranına sahip T100\footnote{\url{https://tug.tubitak.gov.tr/tr/teleskoplar/t100}} ile  odak oranları f/8 olan 60 cm ayna çaplı IST60\footnote{\url{https://arlab.istanbul.edu.tr/arlab/front/browse_lab.html?labid=185}} ve 50 cm ayna çaplı ATA50\footnote{\url{hhttps://ata50.atauni.edu.tr/}} teleskoplarından alınan veriler sunulmuştur. Ayrıca çalışmamızda kullanın literatür, TESS ve ETD verileri de Tablo \ref{T:obs}’de verilmektedir . ETD’den alınan geçiş zamanlarından çalışmaya dahil edilenler DQ (Data Quality) değerleri 1-3 aralığında olan verilerdir.

\begin{table}
	\centering
	\caption{HAT-P-16b ve TrES-3b UT50 gözlem bilgileri.}
	\begin{tabular}{ l c c c l }
	\hline\hline
	{Ötegezegen} & {Gözlem Tarihi} & {Filtre} & {Poz Süresi (s)} & {Binning}  \\
	\hline
	HAT-P-16b & 09.11.2019 & Rc & 90 & 2x2 \\
	HAT-P-16b & 23.11.2019 & Rc & 60 & 2x2 \\
	HAT-P-16b & 18.12.2019 & Rc & 60 & 2x2 \\
	HAT-P-16b & 21.10.2020 & Rc & 60 & 2x2 \\
	TrES-3b & 17.04.2020 & Clear & 120 & 2x2 \\
	TrES-3b & 17.05.2020 & Clear & 120 & 2x2 \\
	TrES-3b & 21.05.2020 & Clear & 90 & 2x2 \\
	TrES-3b & 07.06.2020 & Clear & 90 & 2x2 \\
	TrES-3b & 24.06.2020 & Clear & 80 & 2x2 \\
	TrES-3b & 03.07.2020 & Clear & 60 & 2x2 \\
	TrES-3b & 28.07.2020 & Clear & 90 & 2x2 \\
	TrES-3b & 04.10.2020 & Clear & 90 & 2x2 \\
	TrES-3b & 21.10.2020 & Clear & 90 & 2x2 \\
	\hline\hline
	\label{T:ut50_obs}
	\end{tabular} 	
\end{table}

\begin{table}
	\centering
	\caption{TESS, ETD, IST60, T100, ATA50 ve önceki çalışmalardan alınan geçiş gözlemi verileri.}
	\begin{tabular}{ l l l }
	\hline\hline
	{Ötegezegen} & {Veri Kaynağı} & {Veri seti Sayısı}  \\
	\hline
	HAT-P-16b & ETD & 36 \\
	HAT-P-16b & ATA50 & 1 \\
	HAT-P-16b & \cite{2010ApJ...720.1118B} & 2 \\
	HAT-P-16b & \cite{2013yCat..35570030C} & 4 \\
	HAT-P-16b & TESS & 7 \\
	TrES-3b & T100 (TUG) & 2 \\
	TrES-3b & IST60 & 1 \\
	TrES-3b & ETD & 182 \\
	TrES-3b &  \cite{2013MNRAS.428..678T} & 7 \\
	TrES-3b &  \cite{Jiang_2013} & 5 \\
	TrES-3b &  \cite{2019...628A.115V}  & 1 \\
	TrES-3b &  \cite{2009ApJ...691.1145S} & 8 \\
	TrES-3b &  \cite{2010MNRAS.408.1494C} & 1 \\
	TrES-3b &  \cite{Kundurthy_2013} & 10 \\
	TrES-3b &  \cite{2013MNRAS.432..944V} & 14 \\
	TrES-3b &  \cite{2011PASJ...63..301L} & 4 \\
	TrES-3b &  \cite{Ricci_2017} & 6 \\
	TrES-3b & TESS & 36 \\
	\hline\hline
	\label{T:obs}
	\end{tabular} 	
\end{table}

 Geçiş tarihleri ETD’den takip edilerek gözlem planı yapılmış ve uygun hava koşullarında gözlemler gerçekleştirilmiştir. Kaynakların gözlenebilirliği ilgili web sayfasındaki gözlenebilirlik araçları\footnote{\url{ http://catserver.ing.iac.es/staralt/}} kullanılarak kontrol edilmiştir. Yapılan gözlemlerde teleskobun doğru kaynağa yöneldiğinden emin olmak için ilk alınan veri astrometry.net \citep{2009PhDT.......235L} yazılımının web sayfasına\footnote{\url{ http://nova.astrometry.net/upload}} yüklenmiş, oluşturulan yeni dosya SAOImage DS9 \citep{2000ascl.soft03002S} programında açılıp, kaynağın sağ açıklık (RA) ve dik açıklık (DEC) değerleri kontrol edilmiştir.

HAT-P-16b ve TrES-3b gözlemlerinin AstroImageJ (AIJ, \citealp{2013ascl.soft09001C}) programı kullanılarak aletsel etkilerden arındırılan (bias-dark-flat düzeltmesi),  Barycentric Julian Date (BJD-TDB) zamanına dönüştürülen, belirlenen bir grup mukayese yıldıza göre diferansiyel parlaklıkları elde edildikten sonra hava kütlesi etkisinden arındırılan ve geçiş dışı parlaklığa normalize edilerek oluşturulan geçiş ışık eğrileri EXOFAST (versiyon 1.7) \citep{2013PASP..125...83E} ile modellenmiştir. 

Geçiş ışık eğrilerini modellemek üzere EXOFAST-v1'in web sürümü\footnote{\url{https://exoplanetarchive.ipac.caltech.edu/cgi-bin/ExoFAST/nph-exofast}} kullanılarak  sisteme ilişkin parametreler (i, Teff , Fe/H, P ve e) için başlangıç değerleri ve belirsizlikleri literatürden (HAT-P-16 \citealp{2010ApJ...720.1118B}, TrES-3 için \citealp{2009ApJ...691.1145S}) alınıp, yine aynı web arayüzü ile sağlanan gözlemsel veri Nelder-Mead optimizasyon tekniğiyle \citep{1988SPIE..880..187D} uyumlandı (fit edildi) ve sonuç olarak bu parametrelerin en optimize değerleri elde edildi. Uyumlamalar sırasında kenar kararma katsayıları serbest bırakıldı, ancak filtre bilgisi girildi. Ayrıca yıldızın sıcaklığı da fit parametreleri arasında olduğundan kenar kararma katsayılarının başlangıç değerleri EXOFAST-v1 tarafından, \cite{2011A&A...529A..75C} çalışmasındaki tablolardan ilgili sıcaklık için ara değer hesabı yapılarak (interpolasyon) belirlendi. Yörünge dönemi, referans yayından çekilen değer başlangıç parametresi yapılarak uyumlanırken, ana amaç olarak belirlenmesi istenen geçiş ortası zamanı başlangıç değeri olmaksızın serbest bırakıldı. Uyumlandırmanın başarısı indirgenmiş ki-kare istatistiği ve RMS (Root Mean Square) bağlamında değerlendirilirken, bulunan parametrelerin ve geçiş ortası zamanının "doğruluğu" ise geçiş süresi ve geçiş derinliğinin literatürle uyumu bağlamında değerlendirildi. Ayrıca EXOFAST sonuç dosyasından elde edilen parametrelerin Microsoft Excel programında  ağırlıklı ortalama ve standart sapmalarının hesaplanmasıyla sistem parametreleri belirlenmiştir. Elde edilen kütle, yarıçap ve yörünge eğimi gibi sistem parametreleri incelenen sistemin keşif makalesindeki (\citealp{2010ApJ...720.1118B, 2007ApJ...663L..37O}) değerlerle karşılaştırılmıştır (Tablo \ref{T:both_res}).
Veriler EXOFAST-v1 ile modellendikten sonra, geçiş ortası zamanları ve belirsizlikleri elde edildi (Aladağ 2021). Gözlenen bu geçiş zamanları (O) ile referans efemeris parametreleri ($T_0$ ve P) kullanılarak hesaplanan geçiş ortası zamanları (C) arasındaki farklar (O-C) hesaplandı. Bu farklara \textbf{\textit{emcee}} paketi fonksiyonlarına \citep{2013PASP..125..306F} dayanan bir Python kodu ile birer doğru ve parabol uyumlaması yapıldı. Uyumlamlanan fonksiyonların katsayıları, en küçük kareler yöntemiyle yapılan uyumlamalardan elde edilen en iyi değerleri merkez, parametre hatalarını ise standart sapma olarak kabul eden normal dağılımlardan rastgele seçildi. Bu seçim için 50 rastgele yürüyüşün (random walk) 5000 adım boyunca parametre uzayını taramasıyla her bir parametre setinin ardıl olasılık dağılımları hesaplandı. Bu dağılımların medyan (ortanca) değeri uyumlanan fonksiyonun ilgili katsayısının değeri, \%16 ve \%84 yüzdelik değerleri ise belirsizliklerinin alt ve üst limitleri olarak belirlendi. 
O-C değerlerinin çevrime karşı çizdirilmesiyle elde edilen TTV diyagramlarındaki olası dönemlilikleri araştırmak üzere Lomb-Scargle (LS) periyodogramları  kullanıldı. Bu amaçla \textit{astropy} \citep{2018zndo...2556700T} paketinin LS modülü \citep{2018ApJS..236...16V} kullanıldı. 

UT50, IST60, T100, ATA50 teleskoplarından ve TESS uydusundan alınan geçiş gözlemi verilerinin analizleri ile birlikte bu analizlerin sonuçları verilmiştir. Geçiş ışık eğrileri modelleri ile birlikte çizdirilmiştir. Sistem parametreleri, geçiş ışık eğrilerinin modellenmesiyle elde edilen parametrelerin ağırlıklı ortalaması alınarak hesaplanmıştır ve önceki çalışmalardaki değerleri ile karşılaştırmalı olarak sunulmuştur.

\subsection{HAT-P-16b Ötegezegeninin Geçiş Işık Eğrileri}
\begin{figure}
	\begin{center}
		\includegraphics[width=\columnwidth]{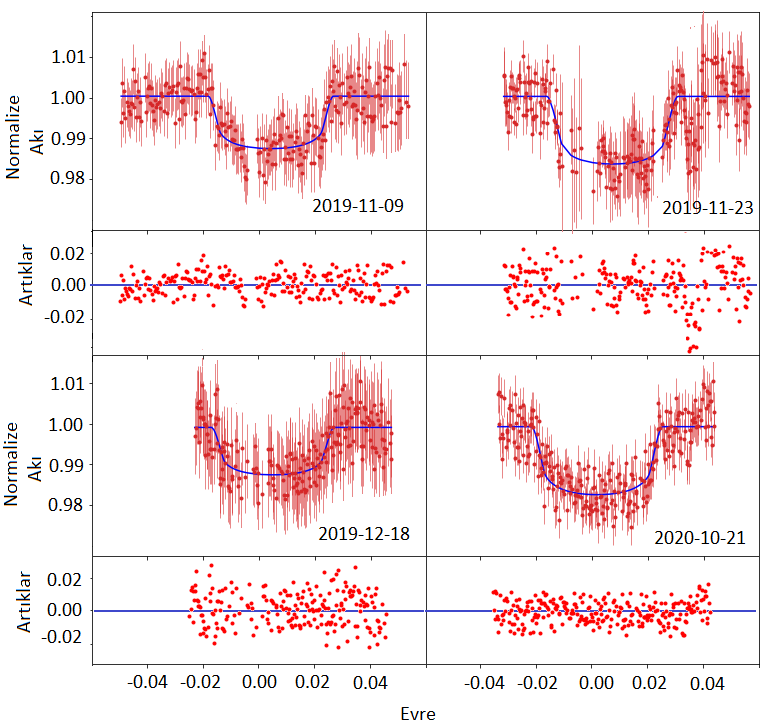}
		\caption{HAT-P-16b’nin UT50 geçiş gözlemleriyle elde edilen ışık eğrileri, hataları, modelleri ve artıkları. Hepsinde geçiş dışı göreli akı 1'e normalizedir. Orjinal veriler ve artıkları kırmızı, modeller mavi renk ile gösterilmiştir.}
		\label{F:h16}
	\end{center}
\end{figure}

\begin{figure}
	\begin{center}
		\includegraphics[width=\columnwidth]{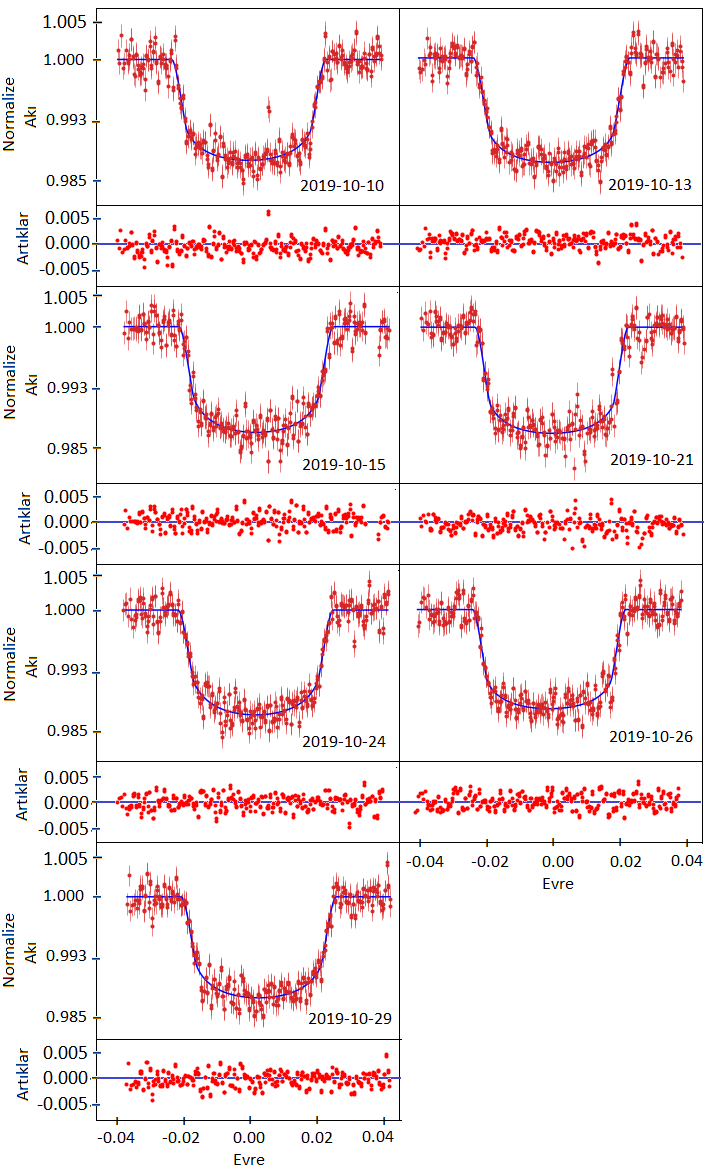}
		\caption{HAT-P-16b’nin TESS ışık eğrileri, hataları, modelleri ve artıkları (Sektör 17). Hepsinde geçiş dışı göreli akı 1'e normalizedir. Orjinal veriler ve artıkları kırmızı, modeller mavi renk ile gösterilmiştir.}
		\label{F:htesss}
	\end{center}
\end{figure}

HAT-P-16b ötegezegeninin geçiş ışık eğrileri yörünge dönemine evrelendirilerek Şekil \ref{F:h16}'de verilmiştir. TESS gözlem verileri 120 s poz süresinde 8 Ekim 2019 tarihinde alınmaya başlanmıştır. Web sayfasından\footnote{\url{https://exo.mast.stsci.edu/}.} indirilen, TESS yazılımı tarafından (pipeline) kırmızı gürültüden arındırılmış ışık eğrileri (dvt uzantılı dosyalardaki LC-DETREND sütunundaki akılar) yaklaşık 27 günlük gözlemleri içermektedir (Sektör-17). Geçiş öncesini ve geçiş sonrasını içerecek şekilde evre aralığı seçilerek oluşturulan ışık eğrileri Şekil \ref{F:htesss}'de verilmiştir.

\subsection{TrES-3b Ötegezegeninin Geçiş Işık Eğrileri}
\begin{figure}
	\begin{center}
		\includegraphics[width=\columnwidth]{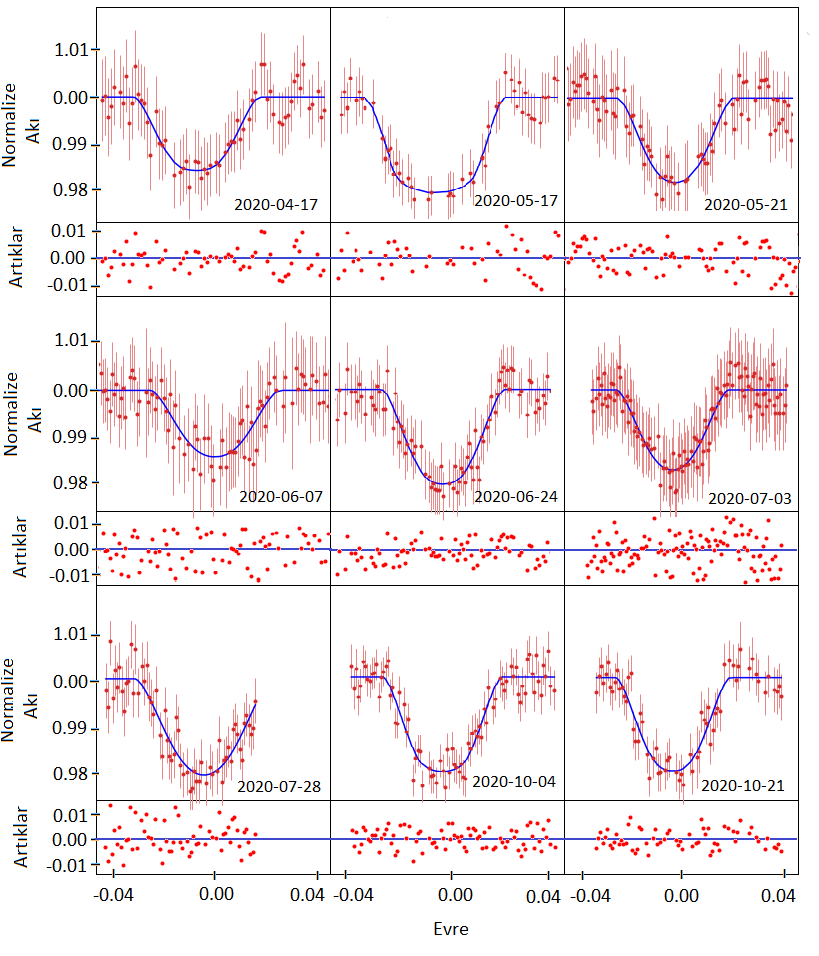}
		\caption{TrES-3b’nin UT50 geçiş ışık eğrileri, hataları, modelleri ve artıkları. Hepsinde geçiş dışı göreli akı 1'e normalizedir. Orjinal veriler ve artıkları kırmızı, modeller mavi renk ile gösterilmiştir.}
		\label{F:tress}
	\end{center}
\end{figure}

\begin{figure*}
	\begin{center}
		\includegraphics[width=\textwidth]{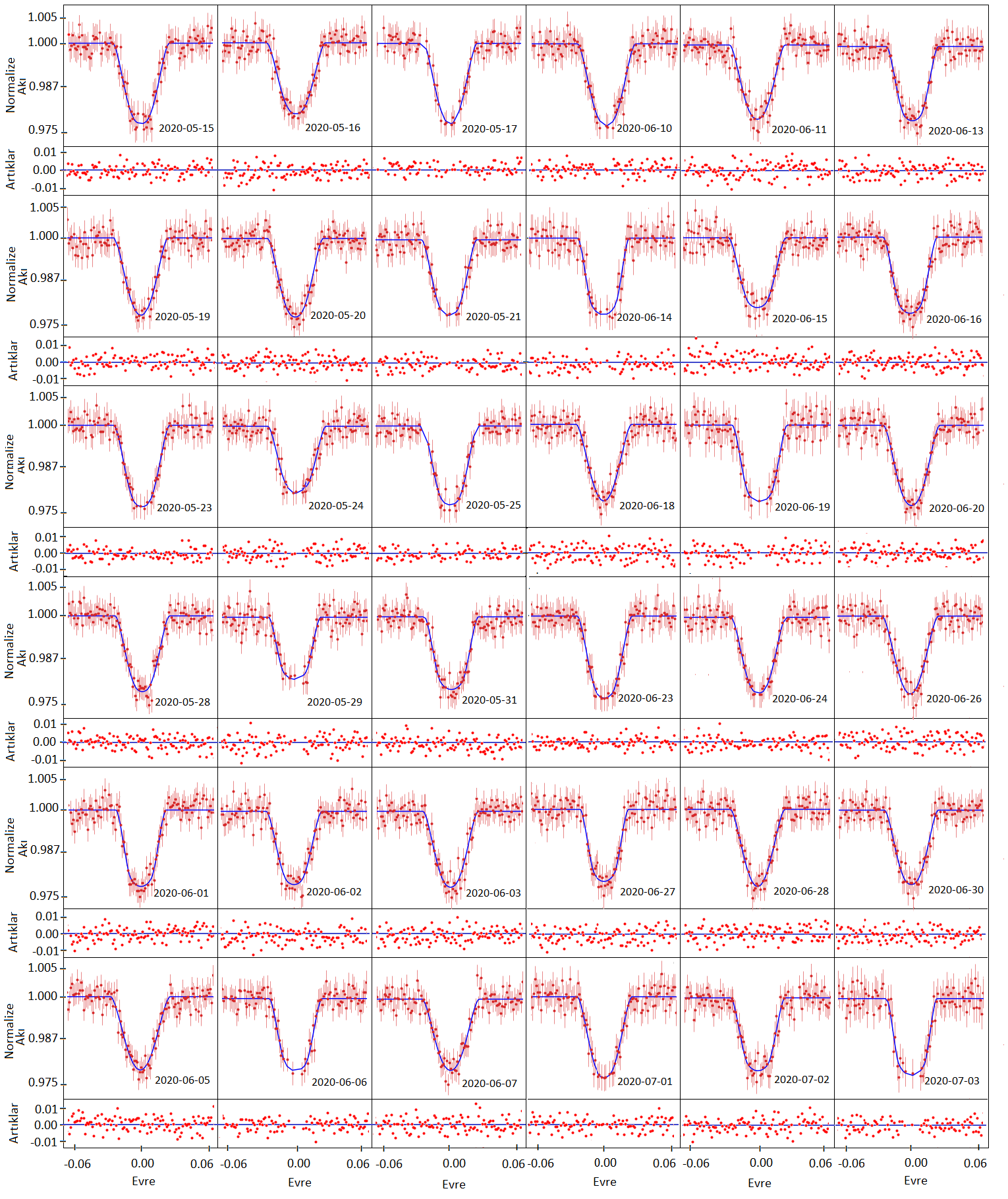}
	         \caption{TrES-3b’nin TESS ışık eğrileri, hataları, modelleri ve artıkları. Işık eğrileri 3 lü paneller olarak sırasıyla sol (Sektör 25) ve sağ (sektör 26) gösterilmiştir. Tüm eğrilerde geçiş dışı göreli akı 1'e normalizedir. Orjinal veriler ve artıkları kırmızı, modeller mavi renk ile gösterilmiştir.}
		\label{F:TTESS25}
	\end{center}
\end{figure*}

TrES-3b ötegezegeninin UT50 geçiş ışık eğrileri yörünge dönemine evrelendirilerek Şekil \ref{F:tress}'te verilmiştir. TESS gözlem verileri 120 s poz süresinde Sektör-25’ten 14 Mayıs 2020 tarihinde, Sektör-26’dan 9 Haziran 2020 tarihinde alınmaya başlanmıştır. Geçiş öncesini ve geçiş sonrasını içerecek şekilde evre aralığı seçilerek oluşturulan ışık eğrileri (Şekil \ref{F:TTESS25}) her sektör için 18 geçiş olmak üzere toplam 36 geçiş içermektedir. 

%\begin{figure}
%	\begin{center}
%		\includegraphics[width=\columnwidth]{images/TrES-TESS-26.png}
%		\caption{TrES-3b’nin TESS ışık eğrileri, hataları, modelleri ve artıkları (Sektör 26). Hepsinde geçiş dışı göreli akı 1'e normalizedir.Orjinal veriler ve artıkları kırmızı, modeller mavi renk ile gösterilmiştir.}
%		\label{F:TTESS26}
%	\end{center}
%\end{figure}

\subsection{HAT-P-16 ve TrES-3 Sistemlerinin Parametreleri}
UT50, T100, IST60 ve ATA50 ile yapılan geçiş gözlemi verileri ile birlikte TESS uydusundan sağlanan verilerin temel analiz adımlarından sonra elde edilen ışık eğrileri EXOFAST ile modellenmiştir \citep{2013PASP..125...83E}. Modellenen verilerden belirlenen sistem parametrelerinin ağırlıklı ortalamaları ve standart sapmaları Excel programında hesaplanarak sistem parametreleri güncellenmiştir. Her iki kaynağın geçiş ışık eğrilerinden elde edilen yıldız ve gezegen parametreleri, önceki çalışmalarda belirlenen parametrelerin hata sınırları içerisinde yer almaktadır. Tablo \ref{T:both_res}’te HAT-P-16b ve TrES-3b sistem parametrelerinin çalışmamızdan elde edilen değerleri ve literatürde verilen değerleri gösterilmektedir. EXOFAST modellemesinde gezegen kütlesi hesaplanmadığı için, hesaplamalarda (\ref{eq1}) denkleminde kütle oranı q değeri;

\begin{equation}
    \begin{split}
    q & = \frac{M_*}{M_P}
    \end{split}
    \label{eq1}
\end{equation}

olarak değişmez kabul edilmiştir. q’yu hesaplamada HAT-P-16 ve TrES-3 sistemleri için sırasıyla \cite{2010ApJ...720.1118B} ve \cite{2008IAUS..249..261S} çalışmalarındaki parametreler referans alınmıştır. q hesapladıktan sonra bu çalışmada hesaplanan yıldız kütlesi (\ref{eq1}) denkleminde yerine yazılarak ötegezegen kütlesi ve (\ref{eq2}) denklemi \citep{1992drea.book.....B} ile hatası hesaplanmıştır. 
$a = b.c$ ya da  $a = \frac {b}{c}$ ise hata;
\begin{equation} 
    \begin{split}
  (\frac {\Delta a} {a})^2 & = (\frac{\Delta b} {b})^2 + (\frac {\Delta c} {c})^2
    \end{split}
    \label{eq2}
\end{equation}
ile hesaplanır.
\begin{table*}
	\caption{HAT-P-16 ve TrES-3 sistem parametrelerinin bu çalışma ve önceki çalışmalardaki değerleri. Kütleçekim odaklama parametresi olarak tanımlanan Safronov sayısı (\ref{eq3}) denklemi ve hatası (\ref{eq2}) ile hesaplanmıştır.}
	\begin{tabular}{lllllll}
	%\begin{tabular}{@{}l@{~}l@{~}|l@{~}|l@{~}l@{~}l@{~}l@{~}}
	\hline\hline
	{Parametre} & {Birimi} & \multicolumn{2}{c}{HAT-P-16} & &  \multicolumn{2}{c}{TrES-3} \\
	\cline{3-4} \cline{6-7} \\
    &	& {Bu Çalışma} & {\cite{2010ApJ...720.1118B}} & & {Bu Çalışma} & {\cite{2007ApJ...663L..37O}} \\
	\hline
    $a/R_*$ 	&	 - 	&	 7.192$\pm$0.002 	&	 7.17$\pm$ 0.28	& &	 6.066$\pm$0.490 	&	 6.06$\pm$0.10 \\
    Etki parametresi ($b$) 	&	 -	&	 0.1002$\pm$0.1533 	&	 0.439$\pm$0.065	& &	 0.789$\pm$0.070 	&	 0.828$\pm$0.010\\
    Delta (geçiş derinliği) 	&	 - 	&	 0.012$\pm$0.002 	& 	 - 	& &	 0.0276$\pm$0.0053 	&	 - \\
    Yıldız yoğunluğu ($\rho$) 	&	 cgs 	&	 0.9141$\pm$0.0007	&	 2.42$\pm$ 0.35	& &	 2.523$\pm$0.667 	&	2.30$\pm$ 0.06$^c$ \\
    Demir bolluğu [Fe/H] 	&	 dex 	&	 0.165$\pm$0.006 	&	 0.17$\pm$0.08	& &	 -0.190$\pm$0.001 	&	-0.19$\pm$0.08$^c$ \\
    Dış merkezlik($e$) 	&	 - 	&	 0.079$\pm$0.027 	&	 0.036$\pm$0.004	& &	 0.0016$\pm$0.0092 	&	 - \\
    Eğim (i) 	&	 Derece 	&	 89.19$\pm$1.21 	&	 86.6$\pm$0.7	& &	 82.28$\pm$1.06 	&	 82.15$\pm$0.21 \\
    Periyod ($P$) 	&	 Gün 	&	 2.7760 Sabit 	&	 2.7760$\pm$3.$10^{-6}$	& &	 1.3062 Sabit 	&	 1.3062$\pm$1.10$^{-5}$ \\
    Yarı-büyük eksen ($a$)	&	 AU 	&	 0.0412$\pm$2.$10^{-5}$ 	&	 0.0413$\pm$ 0.0004	& &	 0.0229$\pm$4.10$^{-5}$ 	&	 0.0226$\pm$ 0.0013 \\
    Yarıçap ($R_p$) 	&	 $R_J$ 	&	 1.31$\pm$0.11 	&	 1.29$\pm$0.07	& &	 1.32$\pm$0.17 	&	 1.30$\pm$0.08 \\
    Kütle ($M_p$) 	&	 $M_J$ 	&	 $4.172^a\pm0.163^b$ 	&	 4.193$\pm$0.094	& &	 1.959$^a\pm0.111^b$ 	&	 1.920$\pm$0.23 \\
    Denge sıcaklığı($T_{eq}$) 	&	Kelvin 	&	 1625$\pm$3	&	 1626$\pm$40	& &	 1626$\pm$61 	&	 1623$\pm$26$^d$ \\ 
    Safronov sayısı ($\Theta$) 	&	 - 	&	 0.221$\pm$0.011 	&	 0.220$\pm$0.011	& &	 0.073$\pm$0.058 &	 0.038$\pm$0.003$^d$ \\
	\hline\hline
	\label{T:both_res}
	\end{tabular}
	\\ \footnotesize{$^a$ ve $^b$ ile gösterilen parametreler (\ref{eq1}) ve (\ref{eq2}) denklemleri ile hesaplanmıştır. Hata yerine hataların standart sapması kullanılmıştır.\\ $^c$ ve $^d$ ile gösterilen parametreler sırasıyla \cite{2009ApJ...691.1145S} ve \cite{2008ApJ...677.1324T}'nın  çalışmalarından alınmıştır. }
\end{table*}

\begin{equation}
    \begin{split}
    \Theta & = (\frac {1} {2})(\frac {v_k} {v_y})^2 = \frac {a M_P} {R_P M_*}
    \end{split}
    \label{eq3}
\end{equation}

Burada, $v_k$; kaçış hızı ve $v_y$; yörünge hızını ifade eder. Denklem (\ref{eq3})'ten Safronov sayısı HAT-P-16b ve TrES-3b için sırasıyla 0.22115 ve 0.07338 olarak hesaplanmıştır.

Ağırlıklı ortalamalardan hesaplanan parametre değerleri, (\ref{eq1}) denkleminden elde edilen $M_P$, HAT-P-16b için \cite{2010ApJ...720.1118B} çalışması, TrES-3b için \cite{2007ApJ...663L..37O} çalışmasında belirtilen değerler ile uyumludur.
{\textquotedblleft}Foton gürültüsü{\textquotedblright}  ya da {\textquotedblleft}Johnson gürültüsü{\textquotedblright}  olarak da adlandırılan beyaz gürültü, CCD’nin algılama empedansının neden olduğu termal gürültüdür \citep{6109069}. Fotometri gürültü oranı (PNR) ile sayısallaştırılan beyaz gürültünün, ışık eğrisinin kalitesini değerlendirmek için hesaplanması gerekir. Bunun yanı sıra model parametreleri gezegen-yıldız yarıçap oranı ve yörünge eğimi, PNR değeri büyüdükçe olumsuz etkilenir. \cite{Fulton_2011} bu gürültüyü şu şekilde tanımlanmıştır:

\begin{equation} 
    \begin{split}
    PNR & = \frac {rms} {\surd \tau}
    \end{split}
    \label{eq4}
\end{equation}

Burada rms, ışık eğrisi artıklarının standart sapması olup dakika başına ortalama pozlama sayısıdır (okuma süresi dâhil), $\tau$ (tau) parametresi ise geçişin başlama zamanın bitiş zamanına bölümünü ifade eder. Bu çalışmada PNR Microsoft Excel programında denklem (\ref{eq4}) ile hesaplanmıştır. 
Beta parametresi ile tanımlanan kırmızı gürültü, atmosferik koşullar, hava kütlesi, astrofizik kaynaklı yıldız değişkenliği ve teleskop takibi gibi sistematik hataların neden olduğu gürültüdür. Kırmızı gürültü geçiş derinliğini (dolayısıyla yıldız-gezegen yarıçap oranını) etkiler. Kırmızı gürültü;

\begin{equation} 
    \begin{split}
    \beta & = \frac {\sigma _r} {\sigma _N}
    \end{split}
    \label{eq5}
\end{equation}

formülü ile hesaplanır \citep{2008ApJ...675.1531W}. Burada $\sigma_r$, artıkların N noktasının M bin’e ortalaması alınarak ve binlenmiş artıkların standart sapmasının hesaplanmasıyla elde edilir. $\sigma_N$ beklenen standart sapmadır. Kırmızı gürültünün yokluğunda $\beta=1$ olduğundan, hemen hemen tüm geçişlerde beyaz gürültü baskın olur \citep{2015MNRAS.446.1389P}. Bu çalışmada $\beta$ değerleri Python programında yazılan kodlar ile hesaplanmıştır. Geçiş ışık eğrilerinin kalitesinin değerlendirilmesinde $\beta$ değeri 2.5 üzerinde olan veriler çalışmadan çıkarılmıştır.

\section{Sonuçlar ve Öneriler}
\begin{figure}
	\begin{center}
		\includegraphics[width=\columnwidth]{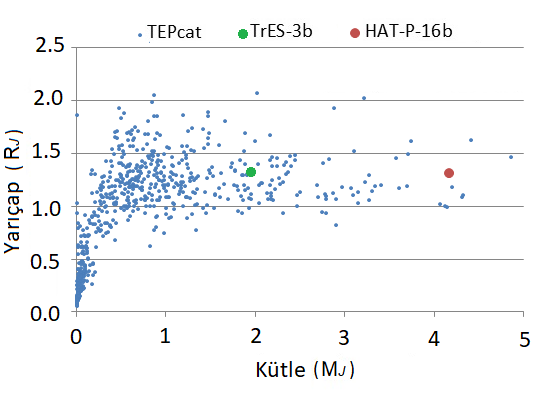}
		\caption{TEPcat verilerinden oluşturulmuş kütle-yarıçap dağılımı}
		\label{F:ytep}
	\end{center}
\end{figure}

Bu çalışmada HAT-P-16b ve TrES-3b ötegezegenlerinin TTV yöntemi ile incelenmesi amaçlanmıştır. Bunun için UT50, T100, IST60 ve ATA50 teleskoplarından alınan geçiş verileri ile birlikte, ETD, TESS ve literatür verileri AIJ programı ile düzenlendikten sonra EXOFAST ile modellenmiştir. Sistem parametreleri hesaplanırken geçiş ışık eğrisi verilerinin ağırlıklı ortalaması kullanılmıştır. Denklem (\ref{eq3}) kullanılarak elde edilen Safronov sayısı, HAT-P-16b için 0.22115 ve TrES-3b için 0.07338 olarak hesaplanmıştır. TrES-3b için hesaplanan bu değer, \cite{2007ApJ...671..861H} tarafından önerilen sınıflamaya göre I. Safronov sınıfını işaret etmekte, HAT-P-16b ise her iki sınıflamaya da girmemektedir. Şekil \ref{F:ytep}.’da  1 Şubat 2021 tarihinde alınan TEPcat\footnote{\url{https://www.astro.keele.ac.uk/jkt/tepcat/}} verilerinden oluşturulan kütle-yarıçap dağılım grafiği gösterilmiştir. Benzer şekilde, \cite{2012MNRAS.426.1291S} çalışmasında ötegezegenlerin Safronov numaralarına göre iki sınıfa ayrılmasının için belirleyici olmadığını göstermiştir. Dahası, \cite{2009A&A...504..605F}’nın oluşturdukları modeller Safronov sayılarının sürekli bir dağılıma sahip olduğunu göstererek bu bulguları desteklemektedir.

\begin{figure}
	\begin{center}
		\includegraphics[width=\columnwidth]{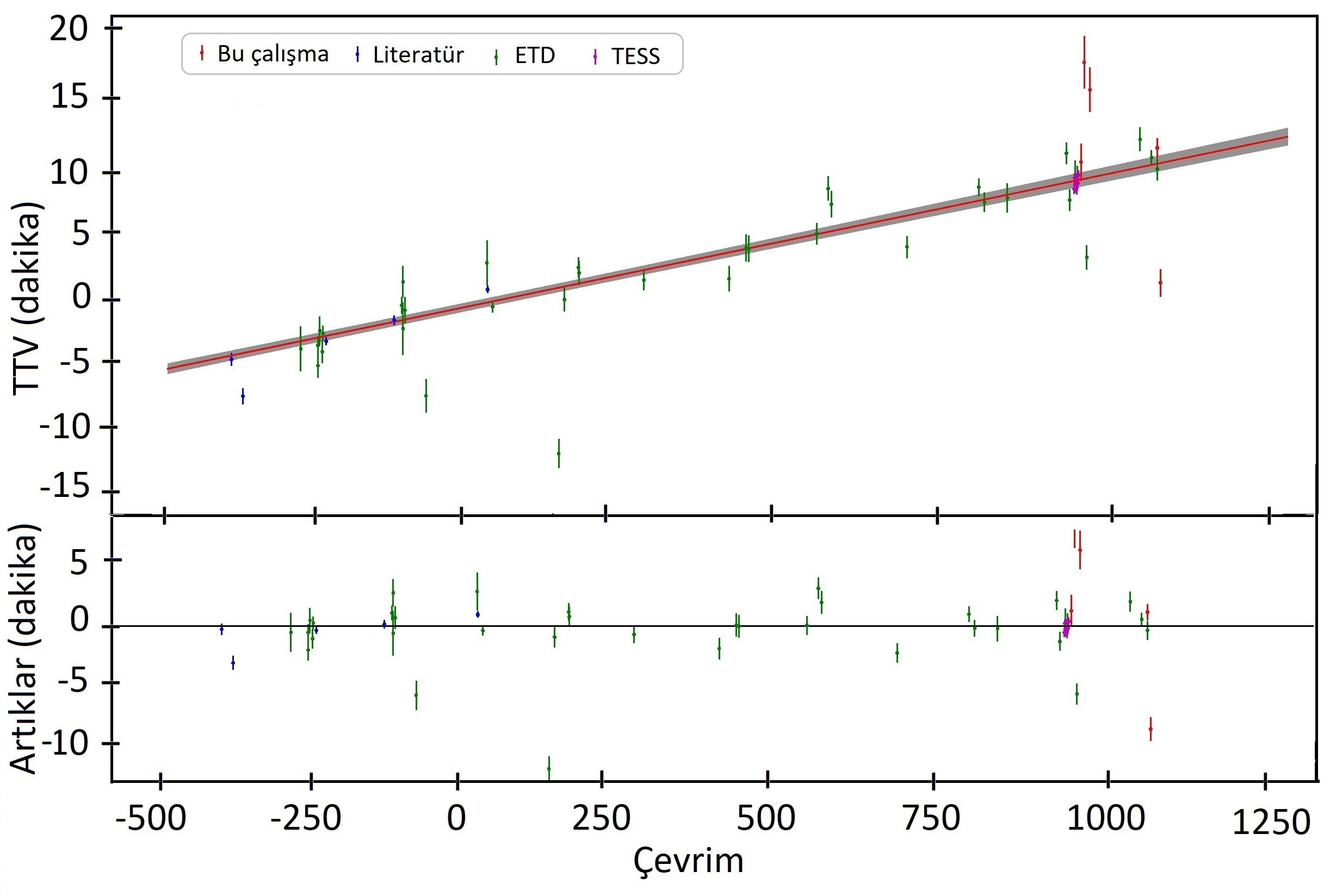}
		\caption{HAT-P-16b'nin 54 geçiş ortası zamanı kullanılarak elde edilen TTV grafiği (üst panel) ve lineer modelden (kırmızı doğru ve belirsizliği gri taralı alan ile gösterilmiştir) artıklar (alt panel)}
		\label{F:hoc}
	\end{center}
\end{figure}

\begin{figure}
	\begin{center}
		\includegraphics[width=\columnwidth]{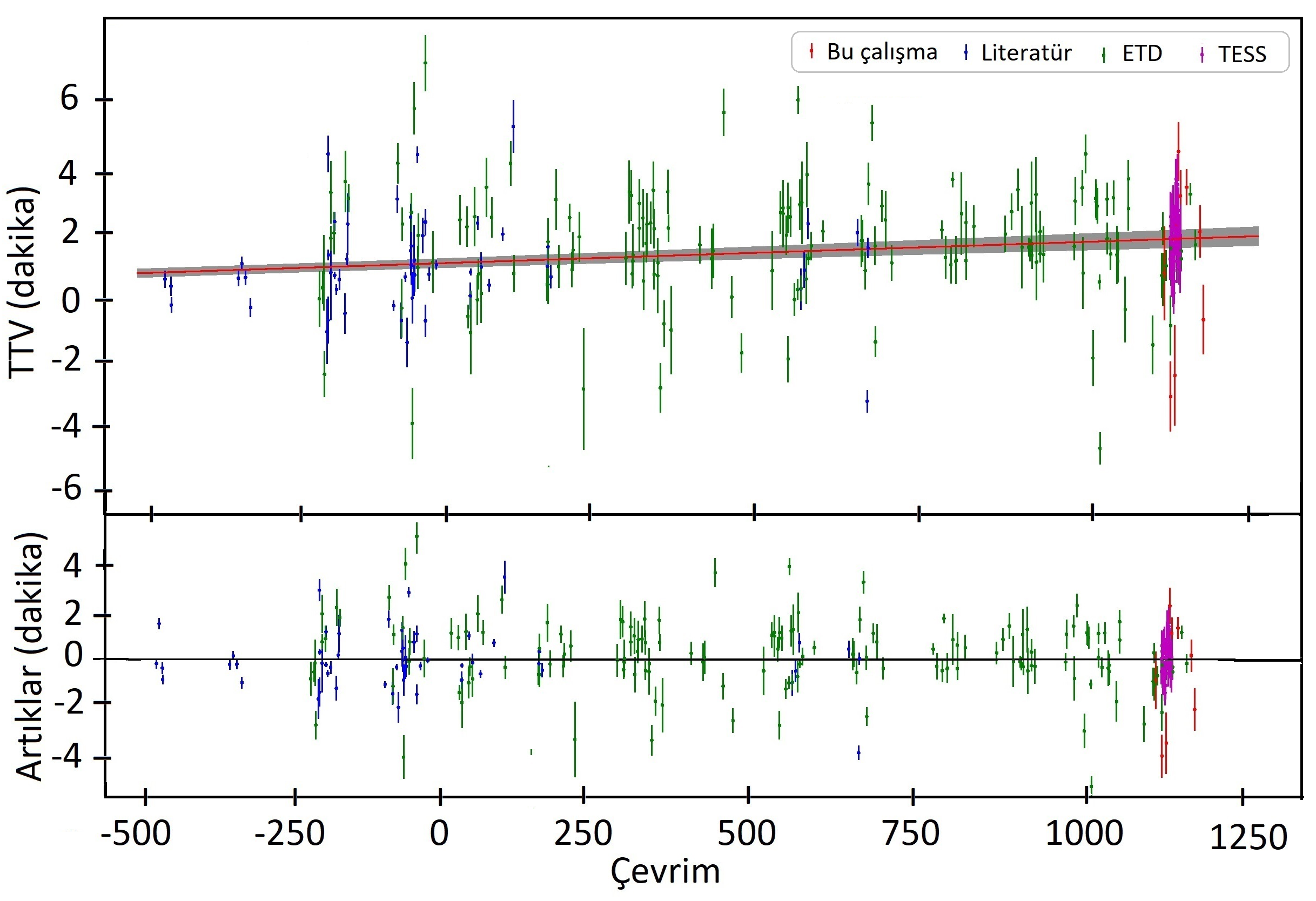}
		\caption{TrES-3b'nin 285 geçiş ortası zamanı kullanılarak elde edilen TTV grafiği (üst panel) ve lineer modelden (kırmızı doğru ve belirsizliği gri taralı alan ile gösterilmiştir) artıklar (alt panel).}
		\label{F:toc}
	\end{center}
\end{figure}

TTV analizleri sonrasında O-C grafikleri elde edilmiştir. O-C hesaplarında, HAT-P-16b’nin $T_0$ değeri ($T_0$=2456204.604209$\pm$0.000318) \cite{2013yCat..35570030C} çalışmasından ve dönemi ETD'den (P=2.77596), TrES-3b’nin $T_0$ değeri ve dönemi ($T_0$=2454552.948971$\pm$0.000147, P=1.30618581$\pm$0.00000051) \cite{2009ApJ...691.1145S} çalışmasından referans olarak alınmıştır. Şekil \ref{F:hoc}’de HAT-P-16b’nin ve Şekil \ref{F:toc}’de TrES-3b’nin TTV grafikleri verilmiştir. 

O-C analizlerinden elde edilen TTV grafiklerine lineer fit uygulanmış, $T_0$ ve P düzeltmesi yapılmıştır. HAT-P-16b‘nin yeni efemerisi;
$T_0$=2456204.604148$\pm$0.000069 ; P=2.77596735$\pm$0.00000011 ve
TrES-3b‘nin yeni efemerisi;
$T_0$=2454552.949831$\pm$0.00107; P=1.30618581$\pm$0.00000051 olarak güncellenmiştir.
Tablo \ref{T:5}'te HAT-P-16b'nin, Tablo \ref{T:6}'te TrES-3b'nin geçiş ortası zamanları ve O-C değerleri gösterilmiştir. Veri seti sayısı çok olduğu için tablolarda Türkiye'de yapılan geçiş gözlemi verileri ve TESS verilerine yer verilmiştir. Çalışmada kullanılan tüm veri setlerinin $T_0$ ve O-C değerleri \cite{Aladag2021} çalışmasında verilmiştir.

\begin{table}
    \begin{center}
	\caption{HAT-P-16b geçiş ortası zamanları ve O-C değerleri (Tabloda UT50, ATA50 ve TESS verilerine yer verilmiştir)}
	\begin{tabular}{lccr}
	\hline\hline
	\centering
	{Teleskop} & {$T_0$} & {$T_0$ err} & {O-C} \\
	\hline
	UT50 & 2458797.358617 & 0.001027 & 0.00096642 \\
    UT50 & 2458811.243992 & 0.001467 & 0.00650461 \\
    UT50 & 2458836.226094 & 0.001233 & 0.00490035 \\
    UT50 & 2459144.346886 & 0.000759 & -0.00668483 \\
    ATA50 & 2459130.474612 & 0.000521 & 0.00087798 \\
    TESS & 2458766.821569 & 0.000287 & -0.00044060 \\
    TESS & 2458769.598131 & 0.000263 & 0.00015404 \\
    TESS & 2458772.373761 & 0.000279 & -0.00018332 \\
    TESS & 2458777.925386 & 0.000288 & -0.00049305 \\
    TESS & 2458780.702133 & 0.000263 & 0.00028659 \\
    TESS & 2458783.477647 & 0.000266 & -0.00016677 \\
    TESS & 2458786.254065 & 0.000263 & 0.00028387 \\
	\hline\hline
	\label{T:5}
	\end{tabular}
	\end{center}
\end{table}

\begin{table}
	\centering
	\caption{TrES-3b geçiş ortası zamanları ve O-C değerleri (Tabloda UT50, T100, IST60 ve TESS verilerine yer verilmiştir)}
	\begin{tabular}{ l c c r }
	\hline\hline
	{Teleskop} &\ {$T_0$} &\ {$T_0$ err} &\ {O-C} \\
	\hline
	UT50 & 2458790.226088 & 0.001233 & 0.00096642 \\
    UT50 & 2458957.409090 & 0.000922 & 0.00650461 \\
    UT50 & 2458987.449195 & 0.000684 & 0.00490035 \\
    UT50 & 2458991.366991 & 0.000806 & -0.00668483 \\
    UT50 & 2459008.348329 & 0.000989 & 0.00096642 \\
    UT50 & 2459025.333177 & 0.000578 & 0.00650461 \\
    UT50 & 2459034.475602 & 0.000507 & 0.00490035 \\
    UT50 & 2459059.294972 & 0.000735 & -0.00668483 \\
    UT50 & 2459127.214106 & 0.000514 & 0.00096642 \\
    UT50 & 2459144.192784 & 0.000684 & 0.00650461 \\
    T100 & 2459017.494626 & 0.000226 & 0.00490035 \\
    T100 & 2459064.518055 & 0.000348 & -0.00668483 \\
    IST60 & 2459106.315072 & 0.000576 & 0.00087798 \\
    TESS & 2458984.839284 & 0.000404 & 0.00000052 \\
    TESS & 2458986.145895 & 0.000512 & -0.00174075 \\
    TESS & 2458987.452491 & 0.000491 & -0.00021302 \\
    TESS & 2458988.758491 & 0.000474 & 0.00043270 \\
    TESS & 2458990.064886 & 0.000388 & -0.00061657 \\
    TESS & 2458991.370023 & 0.000558 & 0.00003288 \\
    TESS & 2458992.677261 & 0.000433 & 0.00043861 \\
    TESS & 2458993.983045 & 0.000654 & 0.00052506 \\
    TESS & 2458995.289637 & 0.000406 & -0.00073321 \\
    TESS & 2458997.902096 & 0.000413 & 0.00058051 \\
    TESS & 2458999.207024 & 0.000634 & -0.00110576 \\
    TESS & 2459000.514524 & 0.000444 & -0.00065403 \\
    TESS & 2459001.819024 & 0.000403 & -0.00021531 \\
    TESS & 2459003.125662 & 0.000468 & -0.00087358 \\
    TESS & 2459004.432287 & 0.000423 & 0.00001615 \\
    TESS & 2459005.737815 & 0.000447 & -0.00273213 \\
    TESS & 2459007.044891 & 0.000557 & -0.00016513 \\
    TESS & 2459008.350896 & 0.000450 & 0.00017505 \\
    TESS & 2459010.964055 & 0.000470 & 0.00114978 \\ 
    TESS & 2459012.269795 & 0.000455 & 0.00028650 \\
    TESS & 2459013.576956 & 0.000397 & 0.00015523 \\
    TESS & 2459014.882279 & 0.000539 & 0.00026096 \\
    TESS & 2459016.188334 & 0.000510 & 0.00019296 \\
    TESS & 2459017.494558 & 0.000409 & 0.00104041 \\
    TESS & 2459018.800075 & 0.004290 & 0.00027113 \\
    TESS & 2459020.107778 & 0.000605 & -0.00023841 \\
    TESS & 2459021.413195 & 0.000418 & 0.00169431 \\
    TESS & 2459024.025058 & 0.000394 & 0.00053031 \\
    TESS & 2459025.332013 & 0.000454 & -0.00035823 \\
    TESS & 2459026.637203 & 0.000465 & -0.00019251 \\
    TESS & 2459027.943497 & 0.000424 & 0.00005422 \\
    TESS & 2459029.249849 & 0.000413 & -0.00031205 \\
    TESS & 2459030.556282 & 0.000400 & -0.00065333 \\
    TESS & 2459031.862102 & 0.000496 & 0.00081540 \\ 
    TESS & 2459033.167947 & 0.000437 & 0.00004240 \\
    TESS & 2459034.474829 & 0.000579 & -0.00043142 \\
	\hline\hline
	\label{T:6}
	\end{tabular} 	
\end{table}

\begin{figure}
	\begin{center}
		\includegraphics[width=\columnwidth]{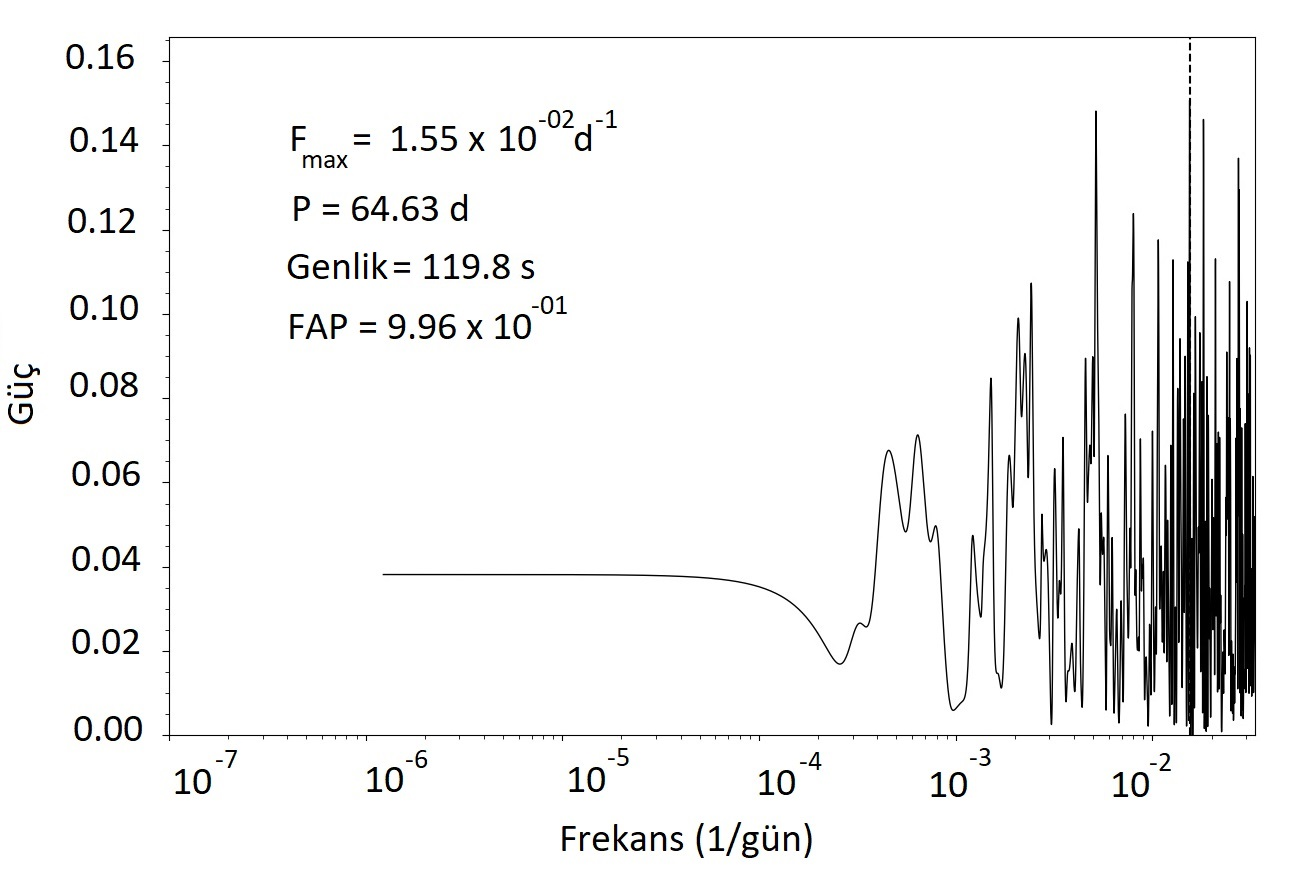}
		\caption{HAT-P-16b’nin LS periyodogramı.}
		\label{F:httv}
	\end{center}
\end{figure}

HAT-P-16b’nin frekans analizinde 54 veri setinden oluşturulan O-C artıkları kullanılarak LS periyodogramı elde edilmiştir. Şekil \ref{F:httv}’da HAT-P-16b’nin LS periyodogramı gösterilmiştir. 

HAT-P-16b’nin frekans analizinde bulunan 64.63 günlük dönem, FAP değeri çok yüksek (\% 99.6) olduğundan istatistiksel olarak anlamlı değildir. Önceki çalışmalarda da belirgin TTV bulgusuna rastlanılmamıştır. Bu kaynak ile ilgili dönem değişiminin izlenebilmesi için daha fazla veriye ihtiyaç vardır. 

\begin{figure}
	\begin{center}
		\includegraphics[width=\columnwidth]{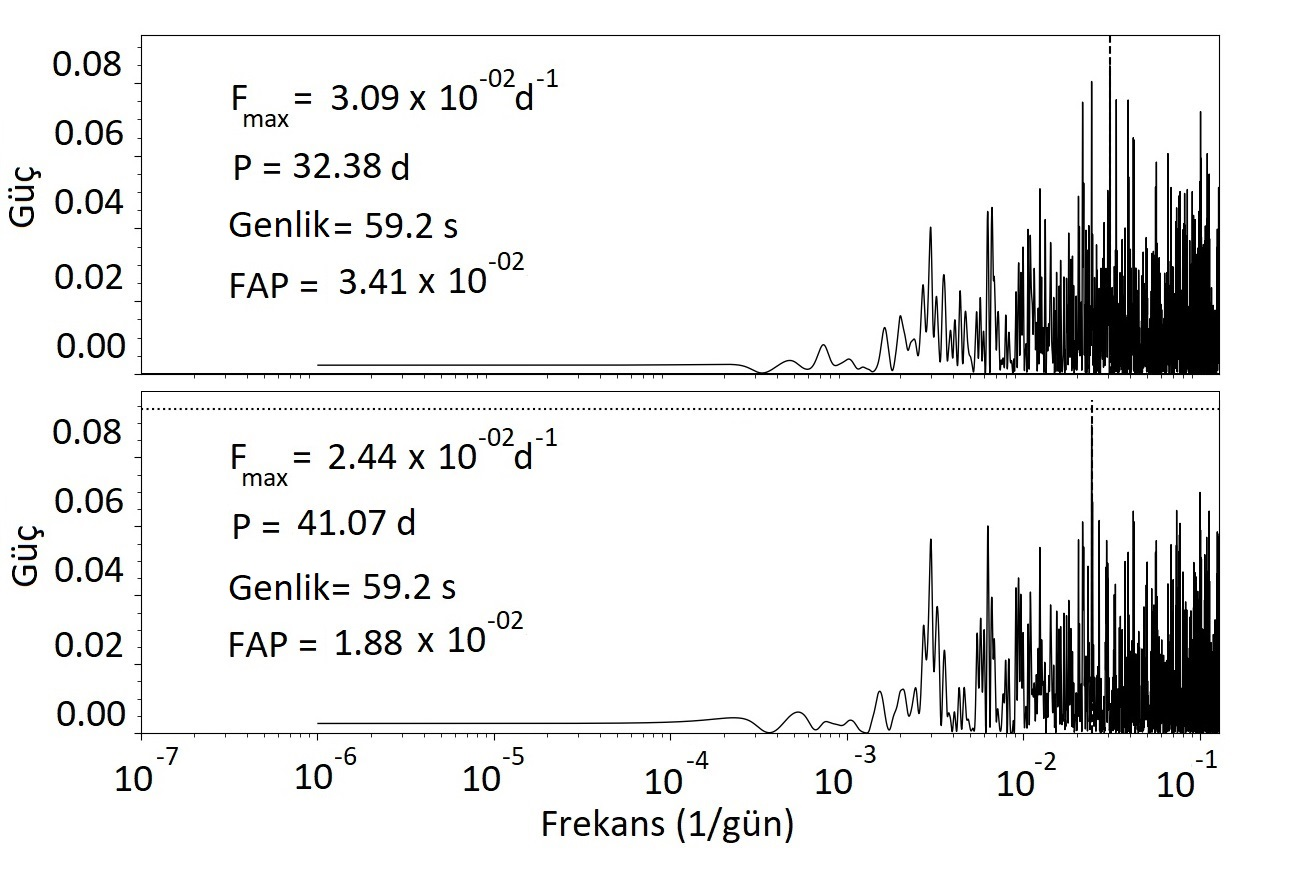}
		\caption{TrES-3b’nin LS periyodogramı.}
		\label{F:tttv}
	\end{center}
\end{figure}

TrES-3b’nin frekans analizinde 285 veri setinden oluşturulan O-C artıkları kullanılarak LS periyodogramı elde edilmiştir. Şekil \ref{F:tttv}’da TrES-3b’nin LS periyodogramı gösterilmiştir.

\begin{figure}
	\begin{center}
		\includegraphics[width=\columnwidth]{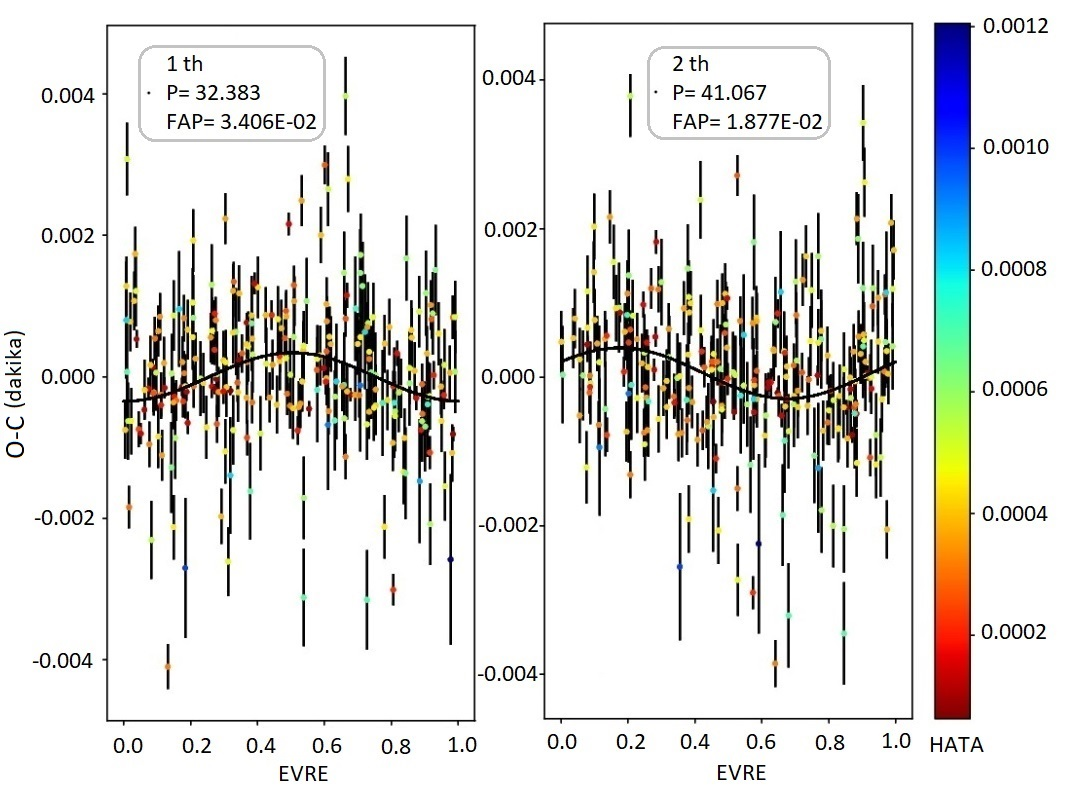}
		\caption{Bu çalışmada elde edilen birincil (32.38 gün) ve ikincil (41.07 gün) döneme göre evrelendirilmiş O-C grafiği (noktaların rengi sahip oldukları hatanın büyüklüğüne göre mavi ile kırmızı arasında değişmektedir).}
		\label{F:evre}
	\end{center}
\end{figure}

TrES-3b'nin frekans analizinde elde edilen LS periyodogramı sonucunda 32.38 günlük bir baskın dönem (FAP=\% 3.41) ve 41.07 günlük ikincil dönem (FAP=\% 1.88) tespit edilmiştir. Şekil \ref{F:evre}’de her iki döneme göre evrelendirilmiş O-C grafiği verilmiştir.

Bu iki dönemden ilki \cite{2020AJ....160...47M} çalışmasında verilen 29.78 günlük değişime çok yakın olup bu değişim olası lekenin dönme modülasyonundan kaynaklı olduğuna işaret etmektedir. Bu değişim, soğuk yıldız lekelerinin geçiş ışık eğrisinde oluşturdukları asimetrilerden dolayı geçiş ortası zamanının doğru belirlenememesine neden olabilir. 
Barınak yıldız kabaca Güneş benzeri (T=5720$\pm$150 K) ve 3 milyar yıl yaşında bir anakol yıldızıdır \citep{2007ApJ...663L..37O}. Bu tür yıldızlarda manyetik aktivitenin neden olduğu böyle bir dönme döneminin beklendiği farklı çalışmalarda sunulmuştur \citep{2010A&A...512A..38L}. Leke sinyali, geçiş eğrileri üzerinde görülmese de neden oldukları asimetriler bu dönemi doğurmuş olabilir. Bunun bir üçüncü bileşen kaynaklı sinyal olduğunu iddia etmek için elimizde yeterince veri bulunmamaktadır.

Çift yıldızların dönem değişimi çalışmalarından bilindiği; \cite{2013ApJS..208...16M, 2018A&A...615A..79V} çalışmalarında da görüldüğü üzere O-C bulguları; veri kalitesi, veri sayısı ve minimum zamanı belirleme yöntemine çok bağlıdır. \cite{2020AJ....160...47M} çalışmasında verilerin bir kısmını yeniden modellemeden, yayınlanmış başka çalışmalardan aldıkları $T_0$ değerlerini kullandıkları için \% 26 FAP'la ${\sim}$30 gün civarında dönemlilik bulmuşlardır. Bizim bu dönemliliği TESS verisi kullanarak ve tüm verileri yeniden modelleyerek ${\sim}$32 gün civarında ve ${\sim}$\% 4 FAP'la bulmuş olmamız önemli bir bulgudur.

Şekil \ref{F:evre}’de kullanılan verilerin gözlem hassasiyeti gereği saçılması çok olduğundan bu dönemlere göre evrelendirilmiş grafiklerde değişim çok belirgin görülmemektedir. Ancak birinci dönem (32.38 gün) değişimin yıldızın dönmesi kaynaklı olduğunu düşünmekteyiz. Bunun nedeni, \cite{2013MNRAS.432..944V} çalışmalarında bu yıldızın dönme dönemi ${\sim}$28 gün olarak belirlenmiştir. İkincisi (41.07 gün) hakkında yorum yapmak için henüz erkendir ve sinyal istatistiksel olarak yeterince anlamlı değildir. FAP=\%1.88 değeri, istatistiksel olarak daha anlamlı olan 32.38 günlük dönemlilik çıkarılmadan hesaplandığı için yanıltıcı olmamalıdır. Her iki dönemlilikten de emin olmak için daha çok veriye ihtiyaç vardır. TrES-3 Haziran ve Temmuz 2021 döneminde TESS tarafından tekrar gözlenecektir ve bu süre içerisinde toplanacak verilerle birlikte sistem incelediğinde bu dönemlilikler konusunda daha anlamlı yorumlar yapılabilir.

\section{Teşekkür}
Bu çalışmada kullanılan Python kodları Doç. Dr. Özgür Baştürk yürütücülüğündeki 118F042 numara ve ''Zamanlama Yöntemiyle Ötegezegen Keşfi'' başlıklı TÜBİTAK 1001 projesi kapsamında geliştirilmiştir. Kendisine bu kodları kullanıma açtığı için teşekkür ederiz. 16BT100-1034 ve 17BT100-1196 numaralı gözlem projeleri kapsamında T100'ün kullanımına izin veren TÜBİTAK'a teşekkür ederiz. ATA50 teleskobu Atatürk Üniversitesi (P.No. BAP-2010/40), CCD kamera ise Erciyes Üniversitesi (P.No. FBA-11-3283) Bilimsel Araştırma Projeleri Koordinatörlüğü Birimi (BAP) tarafından finanse edilmiştir. ATA50'nin kullanımına izin veren Atatürk Üniversitesi Astrofizik Araştırma ve Uygulama Merkezine (ATASAM) teşekkür ederiz. İST60 Teleskobu Türkiye Cumhuriyeti Cumhurbaşkanlığı Strateji ve Bütçe Başkanlığı’nın 2016K12137 numaralı, İstanbul Üniversitesi’nin BAP-3685 ve FBG-2017-23943 numaralı projeleri tarafından desteklenmiştir. İST60'ın kullanımına izin veren Istanbul Üniversitesi Gözlemevi Uygulama ve Arastırma Merkezine (İUGUAM) teşekkür ederiz. UZAYMER'e CCD sağlanmasındaki desteği için TÜBİTAK Ulusal Gözlemevi'ne teşekkür ederiz.

%%%%TJAA-OZEL:BIB%
\bibliographystyle{tjaa}
\bibliography{JTEZ}

%%%%TJAA-OZEL:SON%
\label{lastpage}
\end{document}